# SimAthens: A spatial microsimulation approach to the estimation and analysis of small-area income distributions and poverty rates in Athens, Greece

Anastasia Panori[a,*], Dimitris Ballas[bc], Yannis Psycharis[d]

[a] Department of Economic and Regional Development, Panteion University of Social and Political Sciences, 136 Syngrou Avenue, 17671 Athens, Greece

[b] Department of Geography, University of Sheffield, Winter Street, Sheffield S10 2TN, UK

[c] Department of Geography, University of the Aegean, University Hill, 81100 Mytilene, Greece

[d] Department of Geography, Panteion University of Social and Political Sciences, 136 Syngrou Avenue, 17671 Athens, Greece

* Corresponding author, email: an.panori@panteion.gr

**Abstract**

Published during a severe economic crisis, this study presents the first spatial microsimulation model for the analysis of income inequalities and poverty in Greece. First, we present a brief overview of the method and discuss its potential for the analysis of multidimensional poverty and income inequality in Greece. We then present the *SimAthens* model, based on a combination of small-area demographic and socioeconomic information available from the Greek census of population with data from the European Union Statistics on Income and Living Conditions (EU-SILC). The model is based on an iterative proportional fitting (IPF) algorithm, and is used to reweigh EU-SILC records to fit in small-area descriptions for Athens based on 2001 and 2011 censuses. This is achieved by using demographic and socioeconomic characteristics as constraint variables. Finally, synthesis of the labor market and occupations are chosen as the main variables for externally validating our results, in order to verify the integrity of the model. Results of this external validation process are found to be extremely satisfactory, indicating a high goodness of fit between simulated and real values. Finally, the study presents a number of model outputs, illustrating changes in social and economic geography, during a severe economic crisis, offering a great opportunity for discussing further potential of this model in policy analysis.

# 1. Introduction

Over the past four decades, there have been a rapidly growing number of multidisciplinary efforts to investigate the main aspects of poverty through a wider perspective. In particular, since the 1980s, a new framework has been started to develop, based on the theoretical work of Sen (1983, 1999, 1992), who investigated poverty under the perspective of capability approach. According to this approach, poverty is considered to be caused by not only economic factors but also various components covering wider notions of development such as health, education, and living conditions.

Moving away from conventional poverty measures that include purely income-based indicators and examining poverty under a multidimensional perspective can be challenging due to the paucity of suitable secondary data sets at the small-area level. In order to develop new indicators that attempt to combine and interpret different dimensions of poverty, there is a need for detailed socioeconomic micro-data sets to be collected via social surveys. An example of a comprehensive survey of this type is the census of population, which generates very useful socioeconomic and demographic information for small areas and which has been typically the basis for the development of widely used indexes of deprivation such as the Townsend indicator (Townsend, 1979). Nevertheless, the census questionnaires cover a relatively limited set of themes and in most cases, they do not include any information on income, wealth, and other variables reflecting socioeconomic circumstances in order to preserve confidentiality and minimize nonresponse (Marsh, 1993).

Spatial microsimulation has been gaining prominence as an appropriate method of estimation of small-area microdata that can be used for the analysis of interdependencies between different household and individual characteristics. This method is particularly suitable for bridging the gap between the innovative, human-based theoretical frameworks to examine poverty and the difficulty of implementing it at a small-area level, because of lack of data. Adding geographical information to microlevel data with the use of spatial microsimulation analysis allows for a small-area approach in policy analysis. By using this approach, the distributional impact of implementing different socioeconomic policies could be estimated at a microlevel (Ballas et al. 2005; Ballas et al. 2006; Callan 1991).

Microsimulation models have a long history and tradition in economics, originating in the work of Orcutt (Orcutt 1957) who indicated the importance of determining the relationship between parameters used in a socioeconomic model and the aggregate results. During the 1970s, the first microanalytic models were built to simulate socioeconomic systems and investigate their behavior under various policy implication scenarios (Kain & Apgar 1985; Orcutt et al. 1976).

Although these initial efforts offered a whole new perspective on the way in which aggregate data should be approached, they were aspatial, as they did not include any geographical dimension or perspective. The necessity of incorporating the spatial context was first highlighted by Hägerstrand (1957), who treated time and space as inseparable notions that affect individual's decision making at daily, yearly, or lifetime scales of observation (PRED 1977). The first implementation of such type of model is the work of Wilson and Pownall (1976), which inspired a series of surveys focusing on the field of regional development (Birkin & Clarke 1988, 2011). Other domains in economics where spatial microsimulation has been implemented are labor market (Campbell & Ballas 2013; Ballas et al. 2005; Ballas et al. 2006) and education (Kavroudakis et al. 2012; Kavroudakis & Ballas 2013).

Of particular relevance to the work presented is this study is a comparative study of the social geography of two major cities in Japan and Britain, which involved an estimation of small-area microdata using spatial microsimulation (Ballas et al. 2012). Also of relevance is the work of researchers who developed and implemented statistical small-area estimation approaches involving complementing social survey microdata, such as the European Union Statistics on Income and Living Conditions (EU-SILC) with administrative sources, in order to calculate income and poverty measures based on the idea of utilizing regression models (Fay & Herriot 1979; Elbers et al. 2003; Nagle et al. 2011; Pereira & Coelho 2013; Fabrizi et al. 2014).

The EU-SILC database has proven to be an effective tool, which works as a basis upon which various microsimulation models have been developed. In most cases, the spatial level of analysis remains at a country level, and microdata are used to assess the effect of policy changes, especially referring to national tax benefit systems (Sutherland & Figari 2013; Betti et al. 2010; O'Donoghue et al. 2013).

This study describes the development of the first spatial microsimulation model in Greece that combines census and social survey data, followed by an extensive external validation process, using labor market and occupations structure data. This model aims at estimating small-area poverty measures, including multidimensional poverty index (MPI) within Athens, before and after the economic crisis. An indicative analysis of the main poverty components, as long as the way in which they are affected by recent ongoing economic crisis in Greece, is also performed. This analysis highlights the necessity of using innovative techniques and methodologies for simulating and assessing policy decisions at a microlevel.

**2. Methodology and data**

As mentioned in the previous section, there have only been a handful of studies that attempted to investigate poverty under a multidimensional context at a geographical scale lower than that of countries or NUTS1 regions (Alkire et al. 2014; Miranti et al. 2010; Harding et al. 2006; Tanton et al. 2009). In most cases, the results illustrate that geography matters even at a high level of spatial analysis. There is also a growing literature that refers especially to urban poverty and the importance of structural clusters that are geographically defined within large spatial agglomerations (Amis 1995; Glaeser 1998; Wratten 1995; Satterthwite 1997; Moser 1998). Thus, the need to focus our research on smaller area levels becomes evident, highlighting the necessity of implementing innovative techniques and methodologies.

**2.1 Data, methods and scales of analysis**

In order to build a static spatial microsimulation model, two main sources of data are essential: aggregate data at the spatial level to be used for the analysis and nonspatial microdata. The main idea is to use the existing high-quality aggregate data that have a high degree of accuracy and reliability, such as those derived by national censuses, based on which small-area microdata fitting is acquired, resulting in resynthesized small-area populations.

In this study, the metropolitan area of Athens is used as the main case study. Aggregate data for its 59 municipalities are derived from the last two national censuses (2001 and 2011). The choice of municipalities as the main areal unit is based on the fact that municipality is the lower administrative level at which aggregate data can be found in national censuses. Their size is appropriate to perform spatial analysis in most cases, as they are not too small, leading to a large number of areal units and thus high complexity in calculations. On the contrary, using larger areal units for the analysis could potentially gloss over spatial differentiations within the metropolitan area, reducing its accuracy.

The EU-SILC database was the most appropriate main source of microdata, because of the nature of this research, high quality of the database, and the relative paucity of other relevant survey microdata in Greece. This database contains a large number of parameters referring to economic and social conditions of EU countries. However, although it offers an extremely rich variety of yearly variables that are suitable for poverty analysis, it does not provide geographical information when descending to lower spatial levels of analysis, limiting its use to a country or NUTS1 level.

In order to make a comparative analysis before and after the economic crisis affecting Greece over the past six years, 2006 and 2011 were chosen as the two reference years, and thus the corresponding EU-SILC waves were used. Furthermore, an initial assumption that has to be mentioned is that the 2001 census data were used as aggregate basis for 2006 microdata. It has long been argued (e.g., see Rees, Martin, and Williamson 2002) that when it comes to accuracy and geographical coverage, census data are considered to be the "gold standard." Thus, in order to take advantage of this important strength of census data, it is assumed that between 2001 and 2006, there would have been only small changes in demographic characteristics of the areal units being used here, leaving aggregate data almost unaffected.

The choice of variables being used to constrain a spatial microsimulation model is one of the key factors that play an important role in the process of model building. Different constraint variables may lead to considerable variation in the synthetic populations being produced and thus different results (Edwards et al. 2010; Ballas et al. 2007; Burden and Steel, 2013). The first step when making a decision regarding which small-area variables (known as "small-area constraints") should be used as constraints in spatial microsimulation is to examine the extent to which there is a correlation between these variables with the so-called "target" variables of the simulation (outputs – e.g., income).

Within the context of the research presented in this study, the main variables used for constraining the spatial microsimulation model are age/sex, marital status, education, and main economic activity status. All of them were selected in terms of creating a comprehensive picture of spatial units' demographic characteristics and because they are widely accepted as good indicators of an individual's socioeconomic condition.

**2.2 Spatial microsimulation**

Static spatial microsimulation methods generate simulated microdata for small-area populations at a given time point. Depending on the reweighing method upon which they are based, they can be categorized as either probabilistic or deterministic approaches (Ballas et al., 2005). In the first case, a combinatorial optimization approach (with the use of random number generators) is typically implemented, in order to find the optimal combination of individuals from the micro-data set to reproduce as closely as possible the small-area population (Tanton 2014). The use of a random number generator increases computational intensity of the process (Pritchard & Miller 2011) and a different result is produced in each model run, because of its probabilistic nature.

We choose to follow Lovelace, Ballas, and Watson (2014) and use the so-called deterministic reweighing approach, underpinned by the iterative proportional fitting (IPF) technique as proposed by Ballas et al. (2005). This approach uses the IPF method to give a weight to each individual, by adjusting for each constraint variable the initial weight through a reweighting algorithm (Ballas et al., 2005; Tanton 2014). As a result, individuals whose characteristics match in a higher extent the demographic characteristics of each area are given higher weights. Some of the main advantages of this method include the fact that results remain unchanged in each run of the model, as well as low complexity and high speed of the model (Pritchard & Miller 2011; Lovelace & Ballas 2013; Lovelace & Dumont 2016).

We have also considered alternative approaches for the integerization of the weights derived from the reweighing algorithm. Following Lovelace and Ballas (2013), we choose the TRS (truncate, replicate, sample) method to produce integer micro-data set populations. The main advantage of this integerization method is the fact that the populations it produces have exactly the same size as the census populations, avoiding oversampling. Moreover, Lovelace and Ballas (2013) showed that TRS outperforms the other three investigated methods in terms of accuracy, while at the same time performs really well in terms of speed of calculations.

The original code, based on which the *SimAthens* model was developed, is written by Lovelace and Ballas (2013) and is available as open source, including all the essential documentation (Lovelace and Ballas 2013, Supplementary Information and GitHub - https://github.com/Robinlovelace/IPF-performance-testing). More useful information regarding spatial analysis and microsimulation modeling with R can also be found in Lovelace & Dumont (2016). The new version of the code, used in our case for the development of *SimAthens* model, is also available on line as a supplementary material.

It must be pointed out that the intention of this study is to advance research on census data by combining them with social survey data in new innovative ways, with respect to the investigation of complex socioeconomic phenomena at a regional level, which could not be examined to date. The results are particularly relevant and timely, given the severe crisis and recession affecting Greece over the past 6 years. The spatial microsimulation approach adopted here shows how census data can be combined with social survey data to inform relevant debates about the geographical dimension of income distribution and poverty within Athens.

**2.3 Model validation**

A key step in spatial microsimulation modeling is the verification of model integrity. This is particularly important, in cases where model results are used as inputs for policy-making issues in regional development (Clarke & Holm 1987; Chin & Harding 2007; Smith et al. 2009; Edwards et al. 2010; Ballas et al. 2013).

In order to compare simulation outputs with actual data, there are two main validation methods: internal and external. In the first case, aggregate results of constraint variables are used to calculate the degree of fitting between actual and simulated data. However, because the IPF method is based on optimum fitting of the constraint variables, it is expected that the quality of fitting between actual and simulated results will be high in most cases. The latter method of validating a static spatial microsimulation model consists of using external variables or data sets to compare aggregate results at a regional level (Edwards and Tanton 2013). These external sources of data may be real spatial microdata, external data sets, primary data obtained from specific areas or aggregated data at higher geographies (Lovelace et al. 2014).

The choice of measures of fitting and accuracy is another key decision that needs to be made. We choose to use scatter plots for actual and simulated values, illustrating a first descriptive representation of the model's goodness of fit. In order to calculate the extent of dispersion from the equality line and obtain information about the accuracy of the model, we choose to calculate the standard error about identity (SEI) (Ballas et al. 2007; Tanton 2011). The coefficient of determination ($R^2$) is also estimated through linear regression analysis, as a measure of precision, giving us information about the extent to which simulated values fit the actual data. Finally, an equal variance two-tailed *t*-test is used to reveal any possible statistical significance arising from the differences between simulated and actual data (Robert Tanton; Kimberley L. Edwards 2013).

### 2.3.1 Internal validation

Results of the internal validation of the model, using scatter plots of simulated and actual populations of constraint variables, are shown in **Figure 1**. As it is displayed in the graphs, almost all points of the scatter plots are on the equality line (45° line), showing that the goodness of fit of the *SimAthens* model is excellent for both reference years.

**Figure 1**: Simulated versus census results at the aggregate level for constraint variables.

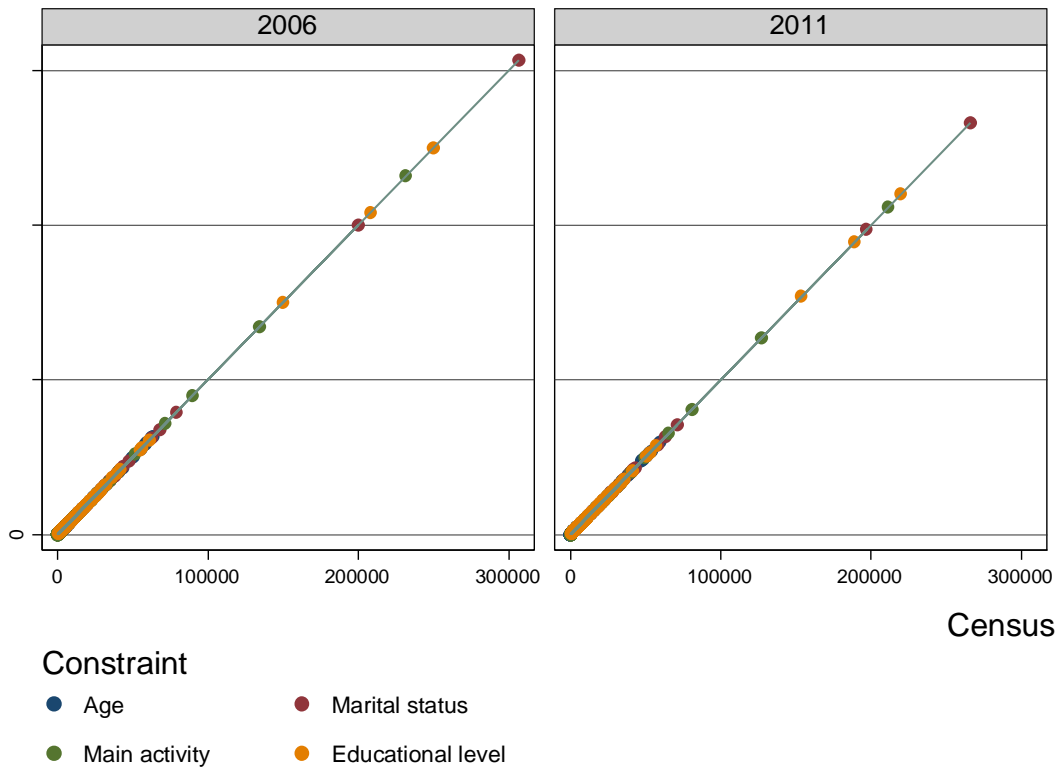

*Source: Authors' calculations*

**Table 1** indicates the values of measures used for internal validation referring to both reference years. It shows that in terms of accuracy and precision, our static spatial microsimulation model shows a very good behavior. Coefficients of determination and SEI values are equal to unity in all cases. The equal-variance two-tailed *t*-test, which was conducted to all subcategories of constraint variables, illustrates that deviations between simulated and actual data are statistically significant in only two cases: Females/20–29 and Females/70–79 for 2011. Thus, estimation results are in general terms robust.

**Table 1:** Measures of validation for constraint variables

| Validation measure | 2006 | | | 2011 | | |
|---|---|---|---|---|---|---|
| | $R^2$ | SEI | T-test (Prob.) | $R^2$ | SEI | T-test (Prob.) |
| **Sex/Age** | | | | | | |
| *Male/20–29* | 1.000 | 1.000 | 0.548 | 1.000 | 1.000 | 0.113 |
| *Male/30–39* | 1.000 | 1.000 | 0.515 | 1.000 | 1.000 | 0.043 |
| *Male/40–49* | 1.000 | 1.000 | 0.087 | 1.000 | 1.000 | 0.126 |
| *Male/50–59* | 1.000 | 1.000 | 0.500 | 1.000 | 1.000 | 0.379 |
| *Male/60–69* | 1.000 | 1.000 | 0.528 | 1.000 | 1.000 | 0.099 |
| *Male/70–79* | 1.000 | 1.000 | 0.705 | 1.000 | 1.000 | 0.017 |
| *Male/80+* | 1.000 | 1.000 | 0.283 | 1.000 | 1.000 | 0.081 |
| *Female/20–29* | 1.000 | 1.000 | 0.442 | 1.000 | 1.000 | 0.008 |
| *Female/30–39* | 1.000 | 1.000 | 0.379 | 1.000 | 1.000 | 0.010 |
| *Female/40–49* | 1.000 | 1.000 | 0.626 | 1.000 | 1.000 | 0.189 |
| *Female/50–59* | 1.000 | 1.000 | 0.851 | 1.000 | 1.000 | 0.206 |
| *Female/60–69* | 1.000 | 1.000 | 0.870 | 1.000 | 1.000 | 0.053 |
| *Female/70–79* | 1.000 | 1.000 | 0.627 | 1.000 | 1.000 | 0.004 |
| *Female/80+* | 1.000 | 1.000 | 0.538 | 1.000 | 1.000 | 0.068 |
| **Marital Status** | | | | | | |
| *Not married* | 1.000 | 1.000 | 0.192 | 1.000 | 1.000 | 0.055 |
| *Married* | 1.000 | 1.000 | 0.810 | 1.000 | 1.000 | 0.065 |
| *Widowed* | 1.000 | 1.000 | 0.080 | 1.000 | 1.000 | 0.060 |
| *Divorced* | 1.000 | 1.000 | 0.541 | 1.000 | 1.000 | 0.596 |
| **Main activity** | | | | | | |
| *Empl./Primary Sect.* | 1.000 | 1.000 | 0.149 | 1.000 | 1.000 | 0.573 |
| *Empl./Second. Sect.* | 1.000 | 1.000 | 0.317 | 1.000 | 1.000 | 0.625 |
| *Empl./Tertiary. Sect.* | 1.000 | 1.000 | 0.828 | 1.000 | 1.000 | 0.436 |
| *Unemployed* | 1.000 | 1.000 | 0.801 | 1.000 | 1.000 | 0.174 |
| *Student* | 1.000 | 1.000 | 0.130 | 1.000 | 1.000 | 0.657 |
| *Retired* | 1.000 | 1.000 | 0.784 | 1.000 | 1.000 | 0.630 |
| *Housework* | 1.000 | 1.000 | 0.881 | 1.000 | 1.000 | 0.930 |
| *Other* | 1.000 | 1.000 | 0.818 | 1.000 | 1.000 | 0.781 |
| **Educational level** | | | | | | |
| *Tertiary* | 1.000 | 1.000 | 0.422 | 1.000 | 1.000 | 0.116 |
| *Secondary* | 1.000 | 1.000 | 0.204 | 1.000 | 1.000 | 0.104 |
| *Primary* | 1.000 | 1.000 | 0.115 | 1.000 | 1.000 | 0.057 |

*Source: Authors' calculations*

### *2.3.2 External validation*

In order to externally validate our static spatial microsimulation model, we chose to use the synthesis of the labor market and occupations. In particular, we used the Statistical Classification of Economic Activities in the European Community (NACE) and the International Standard Classification of Occupations (ISCO) as guidelines for aggregating the actual and simulated data in all areal units. The choice of these classification standards was made, as it was the closest match between the relevant variables found in the EU-SILC and Greek census data.

Although for 2006 all the necessary data for calculating external validation measures at the municipality level (intra-urban municipal units, within Athens) were available, the same was not feasible for the 2011 data. As a result, external validation measures were calculated at the municipality level only for 2006.

*Labor market structure*

**Table 2** illustrates the proportions of aggregated simulated and actual data referring to labor market structure of the metropolitan area of Athens. As it is shown, very small deviations exist between real and estimated values, regarding the proportions of labor market sections. *SimAthens* overestimates in both cases section *G* referring to wholesale and retail trade: repair of motor vehicles and motorcycles. It also overestimates sections from *B* to *E* for 2011 that represent mining and quarrying; manufacturing; electricity, gas, steam, and air conditioning supply and water supply; sewerage, waste management; and remediation activities. Finally, it underestimates section *H*, transportation and storage, for the same year.

**Table 2:** Simulated versus actual census shares for labor market structure in Athens metropolitan area

| 2006 | | | | 2011 | | | |
|---|---|---|---|---|---|---|---|
| NACE Rev 1.1 Sections | Census (%) | SimAthens (%) | Diff. | NACE Rev. 2 Sections | Census (%) | SimAthens (%) | Diff. |
| *A + B* | 0.54 | 0.53 | −0.01 | *A* | 0.66 | 0.53 | −0.13 |
| *C + D + E* | 15.29 | 14.56 | −0.73 | *B − E* | 11.27 | 14.88 | 3.61 |
| *F* | 7.99 | 7.61 | −0.38 | *F* | 6.51 | 7.28 | 0.77 |
| *G* | 18.40 | 21.41 | 3.01 | *G* | 19.04 | 21.96 | 2.92 |
| *H* | 4.98 | 4.09 | −0.89 | *H* | 7.01 | 4.71 | −2.3 |
| *I* | 9.35 | 9.22 | −0.13 | *I* | 5.84 | 4.44 | −1.4 |
| *J* | 4.63 | 3.21 | −1.42 | *K* | 4.30 | 5.52 | 1.22 |
| *K* | 9.15 | 10.64 | 1.49 | *L − N* | 10.74 | 9.92 | −0.82 |
| *L* | 9.80 | 10.42 | 0.62 | *O* | 10.32 | 8.69 | −1.63 |
| *M* | 6.60 | 6.32 | −0.28 | *P* | 7.17 | 7.42 | 0.25 |
| *N* | 5.94 | 5.53 | −0.41 | *Q* | 7.02 | 5.50 | −1.52 |
| *O + P + Q* | 7.32 | 6.46 | −0.86 | *R − U + J* | 10.12 | 9.15 | −0.97 |

*Source: Authors' calculations*

From the scatter plot for labor market structure at municipality level for 2006 in **Figure 3**, we can observe that there is a high goodness of fit between the actual census data and the simulated aggregate results, because of the distribution of scatter plot points very close to equality line.

**Figure 3:** Scatter plot for labor market structure at municipality level (2006).

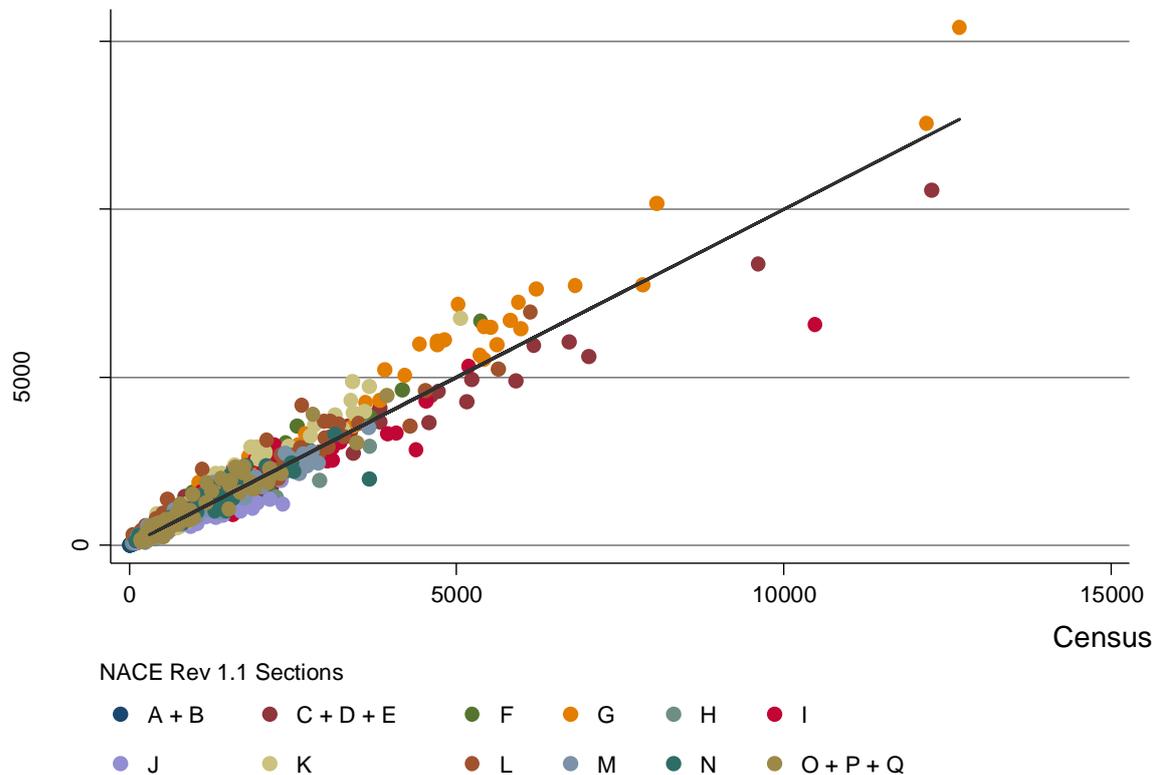

*Source: Authors' calculations*

We can observe from **Table 3** that validation measures are in most cases extremely satisfactory. Coefficients of determination and SEI values are very close to unity in all sections. Moreover, the equal-variance two-tailed *t*-test, conducted in all NACE sections shows that deviations between simulated and actual data are statistically significant in only four sections of the labor market: G (*Wholesale and retail trade; repair of motor vehicles, motorcycles*), J (*Financial intermediation*), K (*Real estate, renting and business activities*) and L (*Public administration and defence, compulsory social security*).

**Table 3**: Measures of validation for labor market structure

| Validation measure | 2006 | | |
|---|---|---|---|
| | $R^2$ | SEI | T-test (Prob.) |
| **NACE Rev. 1.1 Sections** | | | |
| *A + B* | 0.999 | 0.998 | 0.808 |
| *C + D + E* | 0.990 | 0.980 | 0.527 |
| *F* | 0.974 | 0.923 | 0.820 |
| *G* | 0.995 | 0.903 | 0.002 |
| *H* | 0.988 | 0.883 | 0.111 |
| *I* | 0.950 | 0.934 | 0.785 |
| *J* | 0.992 | 0.873 | 0.000 |
| *K* | 0.993 | 0.981 | 0.000 |
| *L* | 0.992 | 0.989 | 0.000 |
| *M* | 0.993 | 0.992 | 0.465 |
| *N* | 0.982 | 0.944 | 0.499 |
| *O + P + Q* | 0.963 | 0.813 | 0.545 |

*Source: Authors' calculations*

***Occupational structure***

The same analysis as before was also conducted in the case of occupations' classification. Using ISCO-88 and ISCO-08 classification system for the two reference years, respectively, we can see in **Table 4** that in both cases there are acceptable deviations between simulated and actual data. Their range varies from 0.02 (*Skilled agricultural and fishery workers,* 2006) up to 4.55 percentage units (*Legislators, senior officials and managers & Professionals.* 2011).

**Table 4**: Simulated versus actual census shares for occupational structure in Athens metropolitan area.

| ISCO - 88 | 2006 | | | 2011 | | |
|---|---|---|---|---|---|---|
| | Census (%) | SimAthens (%) | Diff. | Census (%) | SimAthens (%) | Diff. |
| *Legislators, senior officials and managers & Professionals* | 26.76 | 26.97 | 0.21 | 28.00 | 23.45 | −4.55 |
| *Technicians and associate professionals* | 11.21 | 11.13 | −0.08 | 11.63 | 9.29 | −2.34 |
| *Clerks & Service workers and shop and market sales workers* | 30.25 | 30.42 | 0.17 | 33.23 | 33.45 | 0.22 |
| *Skilled agricultural and fishery workers* | 0.71 | 0.69 | −0.02 | 0.93 | 1.74 | 0.81 |
| *Craft and related trade workers* | 15.61 | 16.87 | 1.26 | 11.46 | 14.46 | 3.00 |
| *Plant and machine operators and assemblers* | 6.96 | 6.25 | −0.71 | 6.07 | 8.08 | 2.01 |
| *Elementary occupations* | 8.50 | 7.67 | −0.83 | 8.68 | 9.53 | 0.85 |

*Source: Authors' calculations*

**Figure 4** clearly shows the high goodness of fit between the static spatial microsimulation model that we have constructed and the actual data for 2006 at a municipality level. All scatter plot points are dispersed around the equality line, indicating a very low level of deviation between the data.

**Figure 4**: Scatter plot for occupation categories at a municipality level (2006).

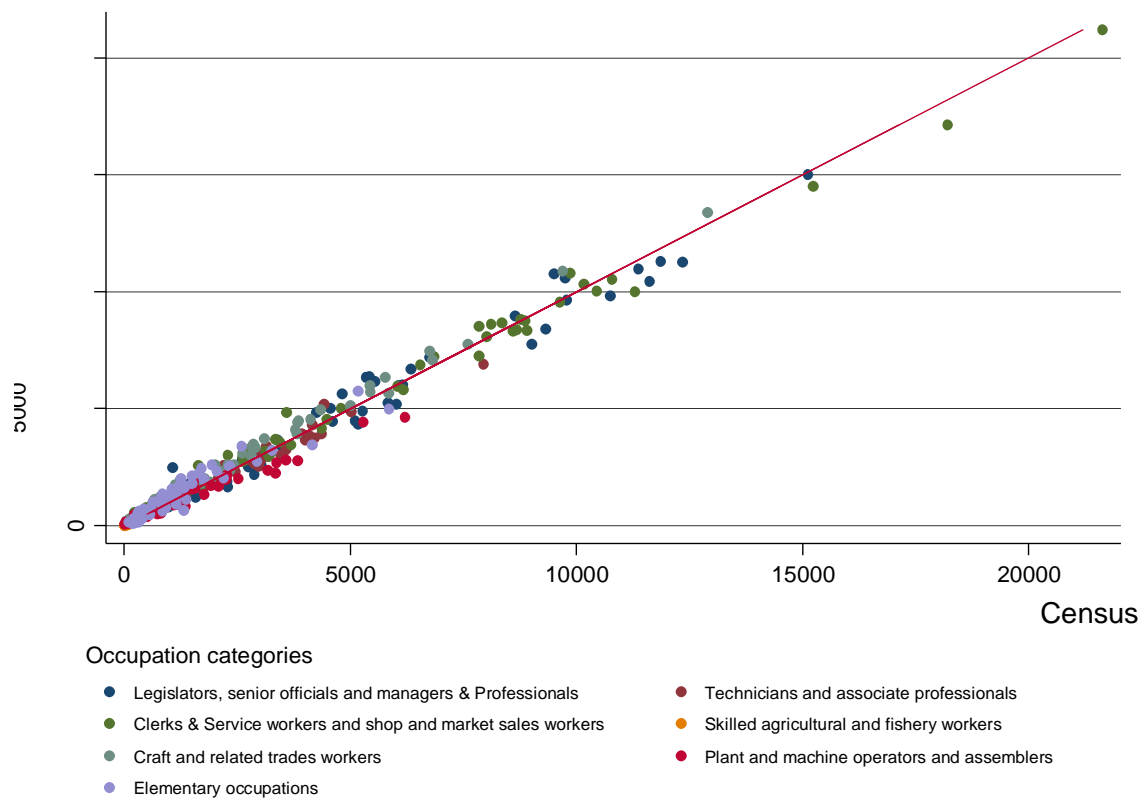

Occupation categories
- Legislators, senior officials and managers & Professionals
- Clerks & Service workers and shop and market sales workers
- Craft and related trades workers
- Elementary occupations
- Technicians and associate professionals
- Skilled agricultural and fishery workers
- Plant and machine operators and assemblers

*Source: Authors' calculations*

The accuracy of the model can also be determined from the validation measures calculated for the same year at a municipality level (**Table 5**). Both $R^2$ and SEI indicators have high values close to unity indicating a good behavior of the model, in terms of accuracy and precision. The equal-variance two-tailed *t*-test was rejected in almost all cases, with the exception of *Craft and related trade workers* and *Plant and machine operators and assemblers* categories, where the deviations seem to be statistically significant.

**Table 5**: Measures of validation for labor market structure

| Validation measure | 2006 | | |
|---|---|---|---|
| | R² | SEI | T-test (Prob.) |
| **ISCO – 88** | | | |
| *Legislators, senior officials and managers & Professionals* | 0.997 | 0.995 | 0.762 |
| *Technicians and associate professionals* | 0.997 | 0.997 | 0.464 |
| *Clerks & Service workers and shop and market sales workers* | 0.999 | 0.997 | 0.851 |
| *Skilled agricultural and fishery workers* | 0.981 | 0.979 | 0.461 |
| *Craft and related trade workers* | 0.998 | 0.995 | 0.000 |
| *Plant and machine operators and assemblers* | 0.983 | 0.972 | 0.002 |
| *Elementary occupations* | 0.960 | 0.790 | 0.541 |

*Source: Authors' calculations*

## 3. Measures of poverty

Over the past four decades, there have been significant efforts to better conceptualize a broader concept of poverty (Townsend 1979; Atkinson 2003; Nolan & Whelan 2011; Alkire et al. 2014). More recently, the European Commission has been contributing to these efforts by adjusting multidimensional measurement of poverty frameworks for advanced economies (Weziak-Bialowolska & Dijkstra 2014; Atkinson & Marlier 2010). The work presented here builds upon these efforts and aims to highlight the regional and local dimensions of the indicators being used to date by the EU, by calculating them at small-area levels. The main measures selected for our analysis are the so-called at-risk-of-poverty (AROP) rate, material deprivation (MD) rate, and MPI. The first two measures consist of simple headcount ratios referring to the proportion of people considered poor each time. However, MPI is based on the Alkire Foster (AF) methodology using an adjusted headcount ratio (Alkire & Foster 2011a; Alkire & Foster 2011b), which includes intensity.

Since 2001, the measure for income poverty in EU has been established by the European Commission as the AROP rate, which is the proportion of people whose income is <60% of the national (in our case municipality) household equivalized median income. Moreover, an indicator concerning *MD* has been proposed and formally agreed by the EU in 2009 (Guio, Fusco, and Marlier 2009; Fusco, Guio, and Marlier 2010; Guio et. al., 2009). This indicator focuses mainly on some key aspects of material living conditions, and is defined as the proportion of people living in households, lacking at least three of a list of nine basic items. In our analysis, we choose to examine the relationship between a relative measure of poverty, using municipality-based thresholds to calculate poverty lines for the AROP rates, and an absolute measure, using MD rates, where the same standard is applied to all regions. Finally, for the calculation of MPI, we used as guideline the recent work of Weziak-Bialowolska and Dijkstra (2014), where the main dimensions of MPI are adjusted at a regional level for the EU regions.

## 4. *SimAthens* results

In this section, we present the results of our simulated small-area populations, where metropolitan area of Athens was used as a case study. The EU-SILC data sets that were used include 3375 (2006) and 2754 (2011) observations for the overall Attica region. It must be indicated that only individuals living in the greater area of Attica were chosen as inputs for the static spatial microsimulation model, instead of using the total number of EU-SILC observations for Greece, in order to increase the accuracy of the results, following a similar approach by Ballas et al. (2005), who used regional subsamples of a national survey in their SimBritain model.

**Table 7** shows the first example of SimAthens outputs: equivalized income. The table shows maximum and minimum estimated values of its distribution within the metropolitan area of Athens. In the last column, their proportional difference before and after the economic crisis is also illustrated. All municipalities experienced a decrease in their mean equivalized income during this period of approximately 9–10%. In particular, for the case of metropolitan area of Athens as a whole, this decrease had a mean value of 9.72%.

**Table 7:** Mean equivalized income per capita and its proportional difference for the five richest and five poorest municipalities of metropolitan area of Athens.

| Municipality | Mean equivalized income 2006 (€ per capita) | Mean equivalized income 2011 (€ per capita) | Difference 06/11 (%) |
|---|---|---|---|
| **Met. Athens** | 14453.32 | 13047.03 | −9.72 |
| **Top 5 by income (2011)** | | | |
| *Psychiko* | 17408.58 | 15766.14 | −9.43 |
| *Filothei* | 17482.81 | 15742.46 | −9.95 |
| *Ekali* | 17334.02 | 15482.58 | −10.68 |
| *Papagou* | 17280.42 | 15440.98 | −10.64 |
| *Neo Psychiko* | 16293.64 | 14798.05 | −9.18 |
| **Bottom 5 by income (2011)** | | | |
| *Keratsini* | 13039.01 | 11615.62 | −10.92 |
| *Drapetsona* | 12921.85 | 11563.58 | −10.51 |
| *Agios Ioannis Rentis* | 12800.99 | 11488.40 | −10.25 |
| *Agia Varvara* | 12803.30 | 11452.38 | −10.55 |
| *Perama* | 12695.54 | 11410.29 | −10.12 |

*Source: Authors' calculations*

On the basis of these estimations, the AROP rates were also calculated for each municipality, to obtain a first income-based estimation of poverty. Two ways of calculating AROP rates were considered, aiming to highlight the difference between *spatially absolute* and *spatially relative* ways of approaching this measure.

In the first case, the poverty line being used to calculate AROP rates is referring to 60% of the median equivalized income of the *total metropolitan area of Athens*. This way of calculating the AROP rate is considered to measure poverty in a spatially absolute way, because all individuals have a common reference line, despite living in different areal units. The results of this measure are illustrated in **Figure 5**, where as we move to more disadvantaged regions, the AROP rate is increasing in both years. Moreover, it is important to note that the proportional difference before and after the crisis also shows an inverse relationship with mean equivalized income. This means that although the headcount ratio of poor people increased in all regions of Athens, the rise was relatively higher in economically depressed municipalities.

**Figure 5:** At-risk-of-poverty rates (%) in absolute terms for municipalities in the metropolitan area of Athens

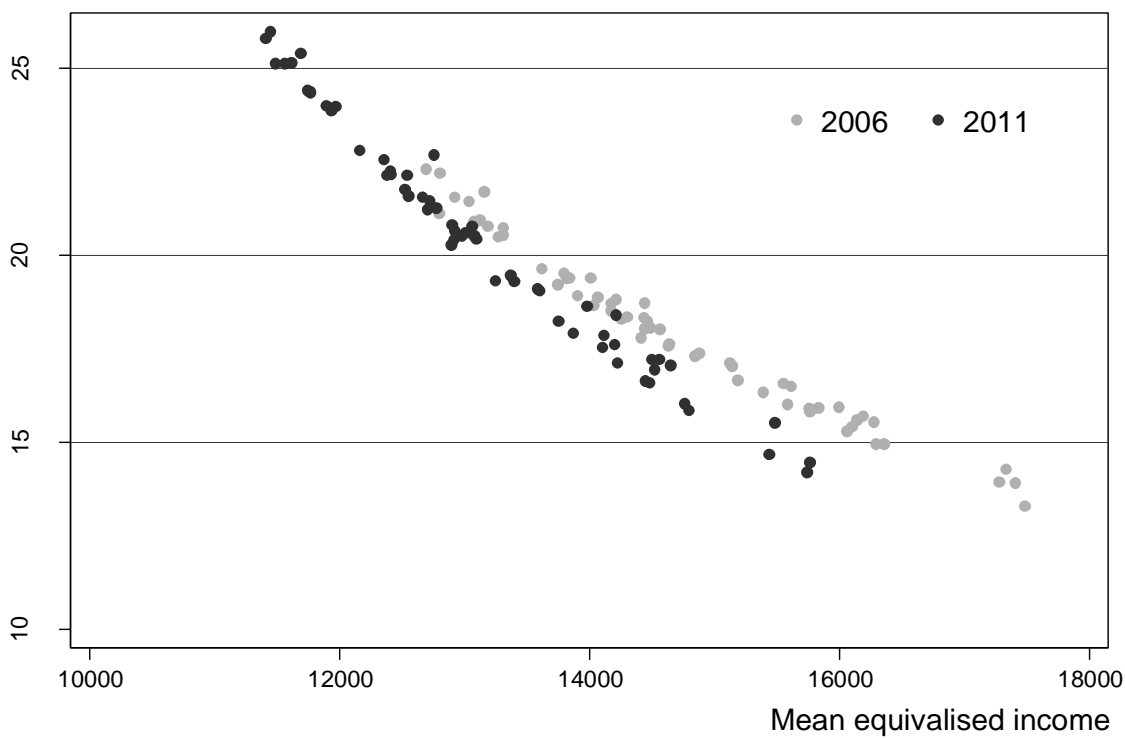

*Source: Authors' calculations*

The previous picture changes radically when using as poverty line the 60% of each municipality's median of equivalized income. In contrast to the previous case, this time each individual is compared only with others living in the same region. As a result, AROP rate in this case has a spatially relative character, exploring inequality within municipal units and using the municipal average income as the point of reference. The results of this measure are given in **Figure 6**. When looking more carefully at the data, it is very interesting to note that in contrast to the previous measure, their values decline while moving to economically depressed areas for 2006. This behaviour could be explained by a possible continuation of the socio-spatial polarization in Athens observed in the 1990s (Maloutas, 2007) underpinned by the tendency of people moving from low income to more affluent areas when they can afford to do so. These interregional movements lead to a more homogenous regional formation, especially for low-income areas, resulting in minor intraregional income inequalities.

Nonetheless, results for 2011 suggest that economic crisis affected primarily low-income areas, where proportions of relatively poor people living in these areas were substantially expanded. By observing the curves illustrated in **Figure 6,** it becomes obvious that spatially relative AROP rates increased in all areas during this period. This indicates that intraregional income inequality expanded, while at the same time gaps in terms of spatially relative AROP rates between areas decreased.

**Figure 6:** At-risk-of-poverty rates (%) in relative terms for municipalities in the metropolitan area of Athens

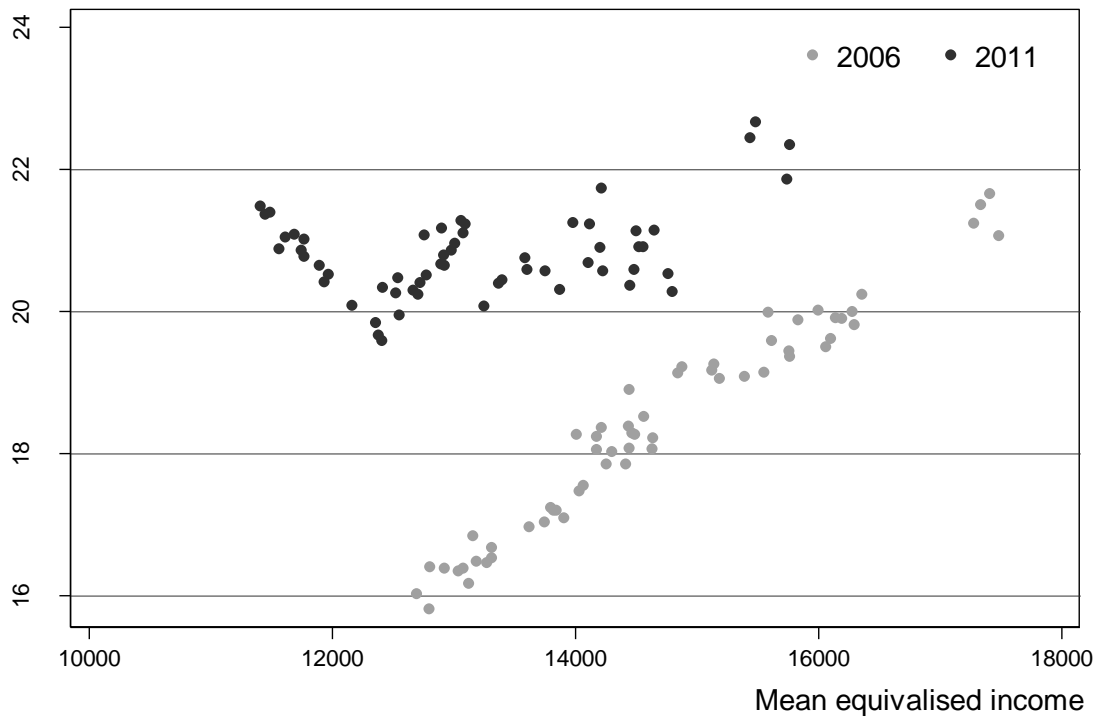

Mean equivalised income

*Source: Authors' calculations*

As a second measure of absolute poverty we choose to use MD rates. An important characteristic of this measure is the fact that MD criteria do not change over time. We can observe in **Figure 7** that MD rates increased during this period, while their values also increased relatively to income levels. Considering the differences between the reference years, it becomes clear that these are much higher than the absolute AROP rates. This is probably due to the time-invariant character of MD rates. Absolute AROP rates may be calculated using a common poverty line for all municipalities, but which still differs between 2006 and 2011. By using MD rates, we can correct this time-dependent inconsistency and this helps us have a clearer view of the crisis effects. Moreover, MD rates offer a more comprehensive view of poverty, by incorporating a diversified view of this phenomenon.

**Figure 7**: Material deprivation rates (%) for municipalities in the metropolitan area of Athens

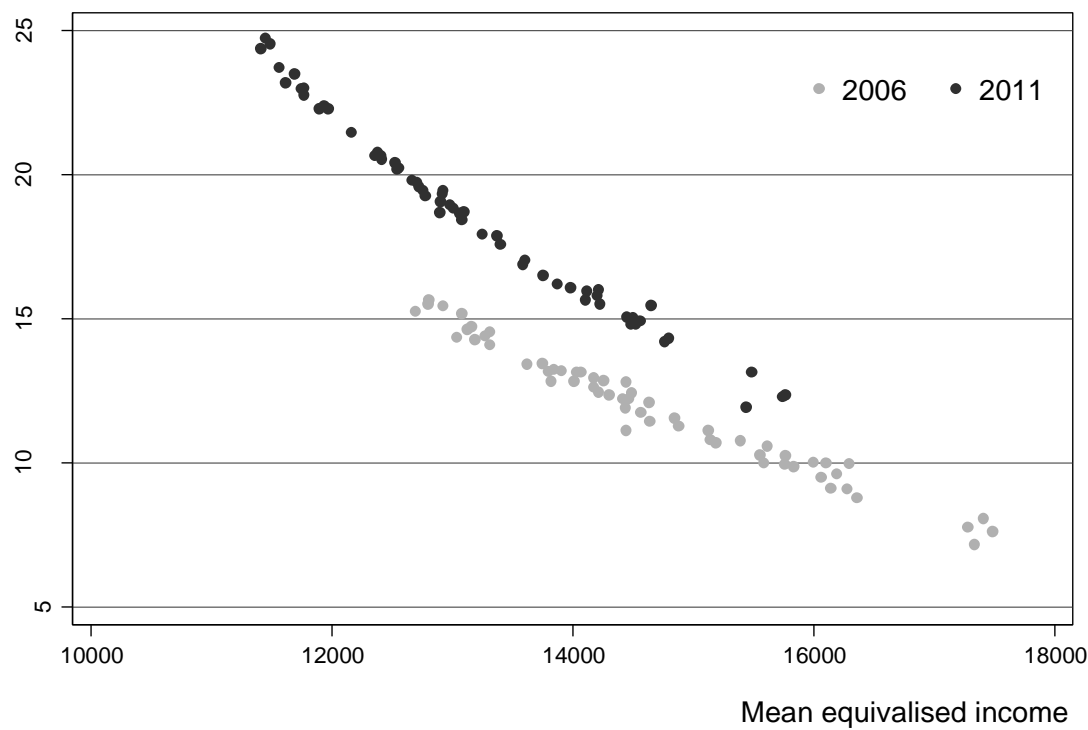

Mean equivalised income

*Source: Authors' calculations*

We have also used the *SimAthens* model to calculate the MPI index at a municipality level and some of these results are shown in **Figure 8**. As can be seen, there is clear evidence that during this prolonged recession period, this measure of poverty increased considerably in most areas.

This observation is of significant interest and relevance to current policy debates, given that the MPI index constitutes a synthetic measure of poverty, not focusing on income, and thus incorporating much information related to aspects of human development and well-being. Another important advantage of this measure is the fact that it is calculated by weighting a simple headcount ratio of multidimensional poor people in each spatial unit, with the intensity of poverty in the same area. As a result, municipalities which are placed at the lowest parts of the ranking in terms of multidimensional poverty perform worse in both headcount ratio and intensity of poverty.

**Figure 8**: Multidimensional poverty index for municipalities in the metropolitan area of Athens

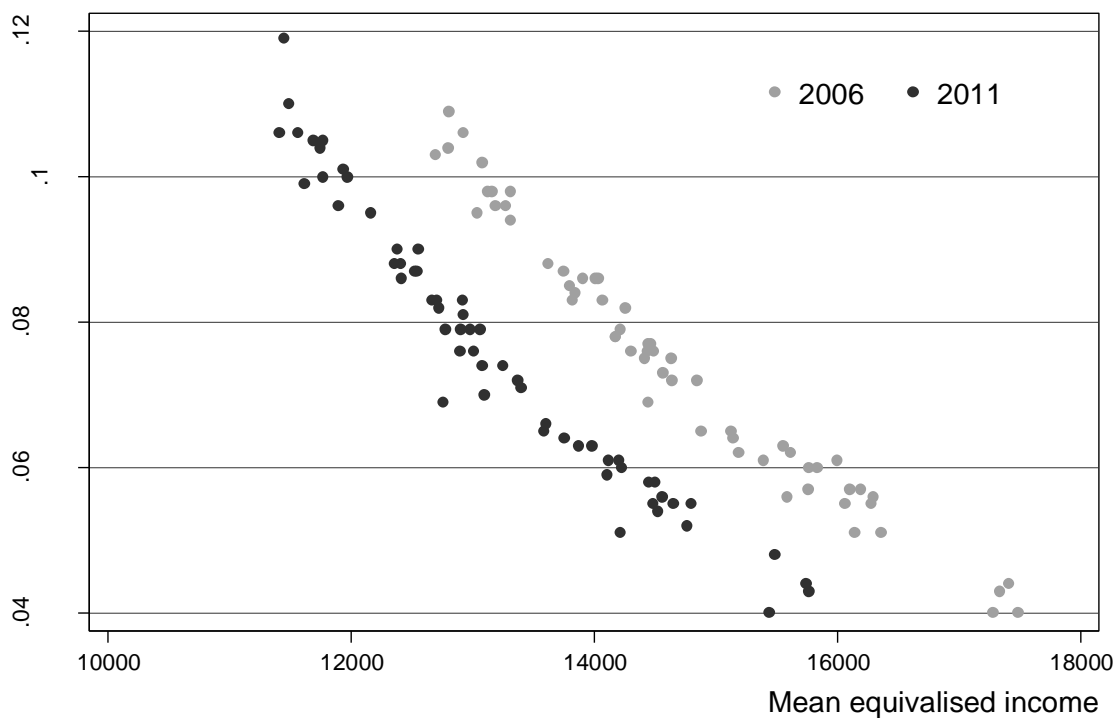

*Source: Authors' calculations*

The results obtained by *SimAthens* model can also help us map and explore the spatial patterns in the distribution of poverty within Athens. **Figure 9** shows the exact location of the metropolitan area of Athens, investigated in this study. Moreover, **Figure 10** depicts the distribution of model aggregate outputs regarding mean equivalized income, MD and AROP rates, within Athens for 2006. As it is illustrated, the more affluent areas are generally located in northeast Athens (**Fig. 10.a**), showing at the same time lower levels of MD and AROP rates in absolute terms (**Fig. 10.b and 10.d**).

**Figure 9:** Location of the metropolitan area of Athens

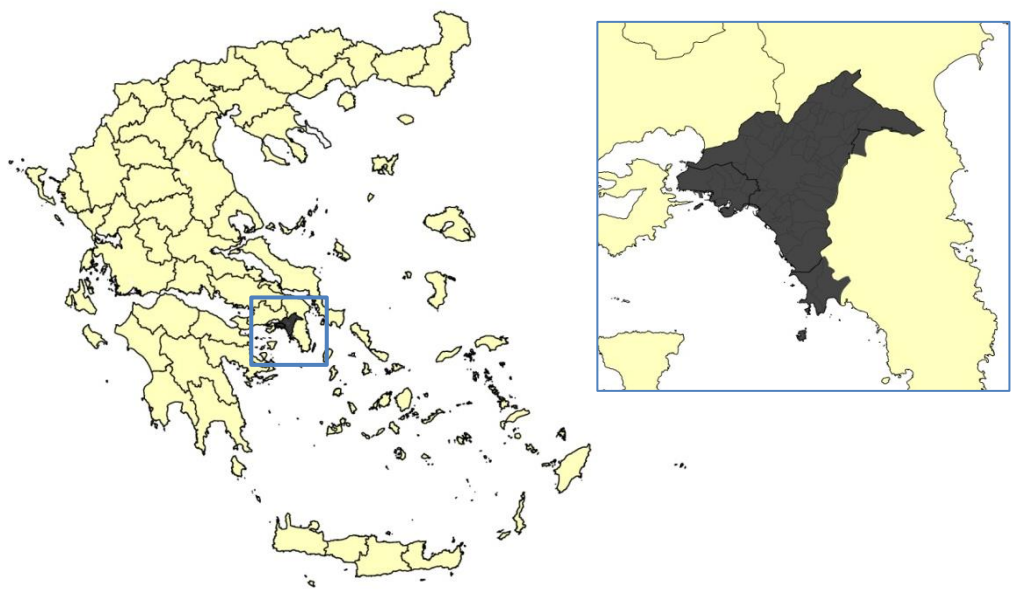

*Source: Authors' calculations*

It is also important to indicate that lower levels of mean equivalized income and poverty are not located in the center of Athens, but that they are generally found in the western part of this large urban agglomeration. However, when concerning the AROP rates in relative terms, these areas seem to have the lowest values, implying low levels of income inequality within these areas. Furthermore, **Figure 10** illustrates that the central part of Athens consists of municipalities that are placed at the middle of income and poverty distribution.

**Figure 10**: Distribution of income and poverty measures within the metropolitan area of Athens (2006).

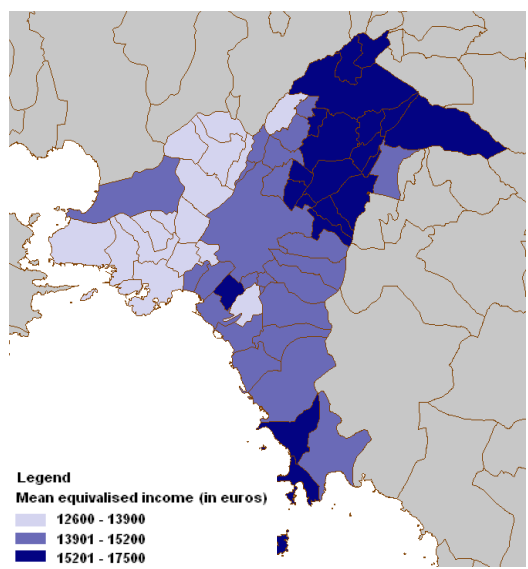

(a) Mean equivalized income

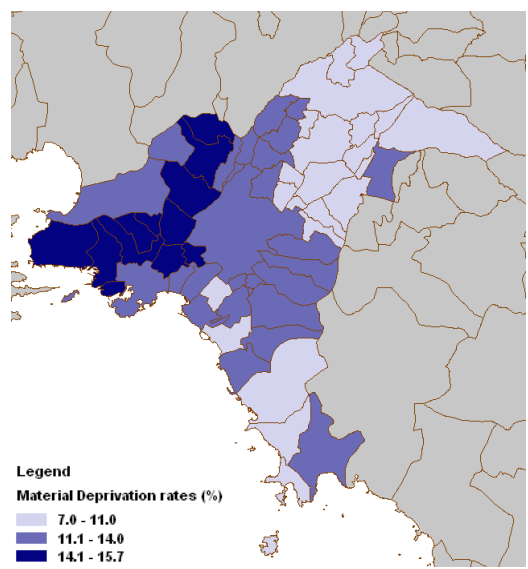

(b) Material Deprivation rate

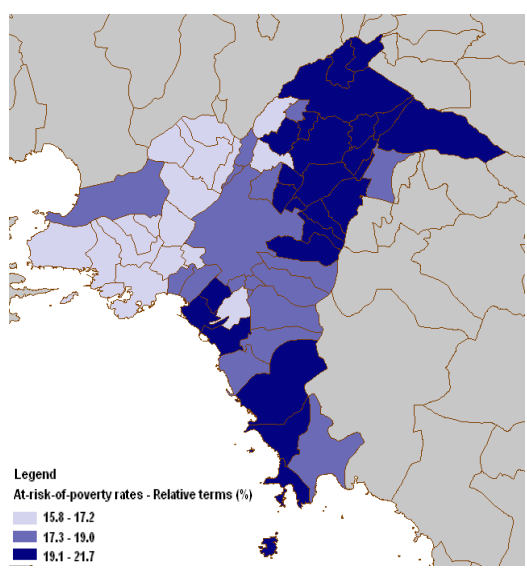

(c) At-risk-of-poverty rate – Relative terms

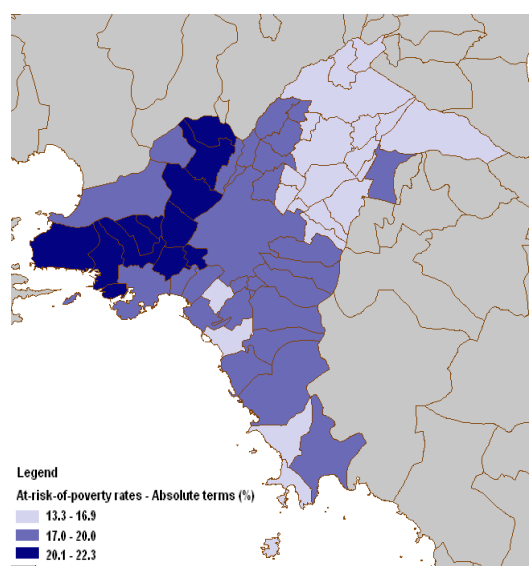

(d) At-risk-of-poverty rate – Absolute terms

*Source: Authors' calculations*

## 5. Concluding remarks

In this study, we have constructed a static spatial microsimulation model that combines small-area census data with social survey data, aiming to analyze income distribution and poverty in Athens. The construction of the *SimAthens* model resulted in the creation of a rich small-area micro-data set for Athens, which was then used for small-area-level analysis of the socioeconomic changes that took place during the period between 2006 and 2011.

After classifying areas based on their mean equivalized income for 2011, the results for the top five and bottom five municipalities were illustrated for both reference years, to investigate the patterns in the two tails of the income distribution. The results indicate that during the period of economic crisis, there was a decrease in all income levels and at the same time a large increase in poverty.

The *SimAthens* model outputs included different measures of poverty focusing on alternative dimensions of poverty and attempting to track differentiations of this phenomenon in space and time using diversified components. Looking at absolute measures, such as the MD and AROP rates in absolute terms, it is interesting to note that their spatial distribution seems to be the converse of the corresponding income allocation. However, because MD rates are time independent (as they do not depend on the average income at the time), there is a higher increase in their values between 2006 and 2011, indicating a relative increase of intermunicipality inequality.

AROP rates in spatially relative terms seem to follow income distribution before the economic crisis, illustrating lower levels of income inequality in poorer regions. However, this situation changed in 2011: relative AROP rates become more homogenous between spatial units of Athens, showing a significant increase in intramunicipality income inequalities for more deprived areas. Finally, looking at the results referring to MPI, we can observe that the values of this measure increase as we move to areas with lower mean equivalized incomes.

It can be argued that the findings presented in this study constitute an initial step of understanding deeper aspects of the ways in which a large city reacts to an economic shock, such as the economic crisis of 2008 and the ongoing recession since then. If the city is assumed a continuously changing dynamic system, the *SimAthens* model presented in this study offers a great opportunity to investigate the underlying dynamics resulting from the individual spatial units composing it. By using the simulated results, to construct either simple or more complex indices, such as the MPI, it becomes obvious that we can expand the analysis to special social groups. Therefore, *SimAthens* could be an appropriate tool to deepen this type of research and investigate possible what-if scenarios for policy-making decisions at a local level.

The research presented in this study provides a glimpse of the potential of combining small-area census data with social survey data, to estimate geographical distributions of variables, for which there are no data available at low spatial levels. To the best of the authors' knowledge, *SimAthens* represents the first effort to build a spatial microsimulation model for Greece. This model could be treated as a platform, on which alternative perspectives regarding the geographical dimension of fiscal consolidation and social policies, as well as the social and spatial impacts of austerity measures in Europe could be tested. For example, at the time of writing this study, the Greek government announced a new round of budget cuts (worth €5.4bn; Smith, 2016) and austerity measures to be introduced in 2016 and 2017, ranging from increases in income tax for particular groups in the labor market (such as the self-employed) to increases in value-added tax and national pension contributions. At the same time, and on a more positive note, there are investment possibilities such as the European Commission *Investment Plan for Europe* (Michalopoulos, 2016), which may offset the impact of austerity to some extent. The geographical as well as socioeconomic impact of all these developments and proposed developments can be analyzed with the use of SimAthens to inform debates and possible help with the formulation of alternative policies and strategies. Overall, *SimAthens* has great potential to be used as a tool for simulating small-area phenomena. Finally, it could provide an effective solution to the analysis of the spatial and socioeconomic impacts of alternative urban, regional, and national social policies, or, in other words, for what-if scenario analysis.